\documentclass[sigplan]{acmart}

\usepackage{booktabs} 

\usepackage{graphicx}
\begin{document}
\title{Blockchain enabled fog structure to provide data security in IoT applications}

\author{Mozhdeh Farhadi}
\affiliation{%
  \institution{U-Hopper, Univ Rennes, Inria, CNRS, IRISA}
}
\email{mozhdeh.farhadi@u-hopper.com}

\author{Daniele Miorandi}
\affiliation{%
  \institution{U-Hopper}
  \city{Trento, Italy} 
}
\email{daniele.miorandi@u-hopper.com}

\author{Guillaume Pierre}
\affiliation{%
  \institution{Univ Rennes, Inria, CNRS, IRISA}
}
\email{guillaume.pierre@irisa.fr}
\renewcommand{\shortauthors}{M. Farhadi}

\begin{abstract}
 IoT provides services by connecting smart devices to the Internet, and exploiting data generated by said devices to enable value-added services to individuals and businesses. In such cases, if data is exposed, tampered or lost, the service would not behave correctly. In this article, we discuss data security in IoT applications across five dimensions: confidentiality, integrity, authenticity, non-repudiation and availability. We discuss how distributed ledger technology could be used  to overcome these issues and propose to use a fog computing architecture as decentralized computational support to deploy the ledger. 
\end{abstract}

%
%



\maketitle

\section{Introduction}
IoT applications, which are based on data captured from the end devices, are vulnerable against a number of different attacks \cite{daniele}. In this paper, we target IoT applications/services that consume data generated by different data providers and are hosted in an environment where there is no trust between the parties, i.e.: data providers, infrastructure providers and IoT applications.
The problem we tackle is the following: how to provide data security for IoT streams generated by different data providers and consumed by different data consumers (service providers via IoT applications) in a trustless environment. Data security has the following five dimensions:
\begin{enumerate}
  \item Confidentiality: The data should not be accessed by any entity other than the source or the legitimate, intended destination of the data. (E.g. : A patient's medical status) 
  \item Integrity: The data received by the receiver should be exactly the data generated by the source. (E.g. : The status of a machine in the production line in Industry 4.0 applications)
  \item Authenticity: The data should be received from the genuine source of data.
  \item Non-repudiation: The data source should not be able to deny it has produced the data
  \item Availability: The generated and stored data should be reliably available.
\end{enumerate}

To meet the aforementioned security requirements, we need computing elements. For example for confidentiality, an encryption element is needed, for non-repudiation, a minimum asymmetric cryptographic support to sign the generated data is needed, etc. On the other hand, one of the characteristics of IoT is heterogeneity: refrigerator, lamp, smartphone or laptop with different computation power and battery life can be part of the network. Thus, for providing security for this heterogeneous network we should add an additional layer on top of the IoT one.

\section{The Approach}
The problem of data security, would be easier to handle if all the generated data belonged to a single administrative entity, but in reality, data may be generated by devices under the control of different administrative entities (`data owners’). A data owner is defined as an administrative entity that provides data (possibly as a service) in an IoT ecosystem. If all the data generated by IoT devices was owned by a single data owner, or data owners trusted each other, then security could be addressed using well-known security techniques. However, what if more than one data owner exist and they do not trust each other? . In \cite{benaloh2009patient}, \cite{damiani2005key}, \cite{zhang2015cloud} and \cite{li2010securing} the problem of data security when there is single or multi data owner is discussed. Most of these works' focus is on data privacy and data confidentiality and they use cryptographic key management solutions to tackle the aforementioned issues. However, in this paper we take a data-centric approach to security issues, and we exploit a decentralized system to address the mentioned data security issues.
Our approach is based on the use of Distributed Ledger Technology (DLT) deployed on a fog infrastructure as data security provider in a multi-data owner and multi-service provider environment, where there is no trust between the mentioned parties. In this paper, with a slight abuse of terminology we will refer interchangeably to DLT as ‘blockchain’.

A blockchain provides an immutable repository in a trustless system \cite{mills}. Blockchain provides a distributed repository, whereby every participating entity has its own copy of the whole ledger; by using a consensus protocol, all the parties agree on a single copy of data to be the correct one at a given time, and this copy of data can not be changed. Blockchains can be either permissionless (anybody can write to the blockchain) or permissioned (only a subset of nodes are allowed to write to the blockchain). Moreover, blockchains can be public (if any entity can verify data in the blockchain) or private (if only participating entity can validate data on the blockchain). 

As in our case, only the devices who generate data or the services that may want to update data should be able to store/change data on the ledger (not attackers or adversaries), we need a permissioned blockchain. On the other hand, as we need that the data be verifiable by all the users/service providers, we propose to use a public permissioned blockchain.
Blockchain is a practically immutable repository, thus integrity of the data is guaranteed by design. Moreover, as it is a distributed system and every node receives and records the data source and the destination, the non-repudiation and authenticity are also supported by design. On the other hand, as the Blockchain is a distributed storage and there will be no single point of failure the problem of availability is also solved implicitly.

\begin{figure}
  \includegraphics[width=\linewidth]{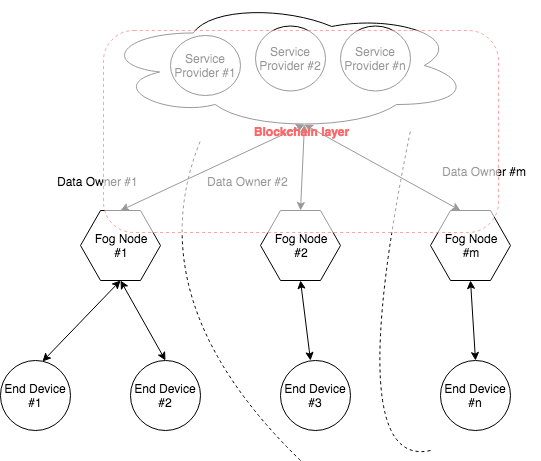}
  \caption{Blockchain layer running on Fog structure}
  \label{fig:boat1}
\end{figure}

To provide data confidentiality, data can be stored as an encrypted data on the Blockchain. The legitimate, intended data receiver, receives a pointer to the location of the data in the Blockchain alongside the proper key to decrypt it. To transfer this key, data source encrypts this key using the public key of the data receiver.
In our proposed architecture, depicted in Figure ~\ref{fig:boat1}, the blockchain acts as a soft layer able to provide data security guarantees for IoT systems. As we need a network of distributed computation elements to deploy the blockchain, the usage of a fog computing approach seems a natural one.

Fog computing is a network of computing nodes that are deployed close to the data sources (closer than cloud) to decrease latency, bandwidth usage and cost in cloud applications \cite{bonomi}. 
The IoT service can be deployed either on the Cloud or on the Fog. As the proposed blockchain is a public, permissioned one, read access to the blockchain is available to everyone, but write access to the blockchain is available only to permissioned parties. Thus, if a data provider, wants to write data on the ledger, it should be defined as a blockchain participant. The consensus protocol that will be used on the blockchain may vary depending on the type of service that we deploy on our architecture. Factors such as acceptable service latency, the computational power of the fog nodes etc. should be taken into account when deciding which specific consensus protocol to use. 
Services that need to access IoT-generated data could read them from any node in the blockchain. In the case of control commands to be sent to IoT devices, the controller should have permissions to write to the blockchain. 
The proposed solution, based upon the usage of DLT, presents of course some costs.  The first is bandwidth usage, because of the fog-to-fog communication required to run the consensus protocol. This also bears a computational cost (related to the processing required to run the consensus protocol). The last cost is storage, as all fog nodes would be required to store the complete ledger, which may grow significantly over time.

\section{The evaluation methodology}
Using attack-tree methodology introduced by Schneier \cite{schneier}, we can analyze the different types of attack and the system reaction. In this graphical approach, the undesirable event will be represented as the root of a tree and using AND and OR gates, various causes that a system can be attacked or defended are analyzed. In attack-tree methodology to assure if the system is robust against attacks, there should be no path from the attacks, which are the leaves of the tree, to the root of the tree. In our case, we will have five different attack-defense trees for each of the mentioned data security dimensions. In the attack-tree approach, blockchain and encryption mechanisms will be represented as a barrier against various sources of attacks, then we should confirm that blockchain layer blocks all the paths to the root of the tree.

\section{Conclusion}
As most of the IoT applications are powered with data generated by end devices, it is instrumental to guarantee security of the data. In this article we use a distributed ledger technology, coupled with a fog computing architecture, to overcome various data security issues.
The author will pursue developing a prototypical implementation of the proposed structure during her PhD studies to analyse attacks and the reaction of the DLT as a defense layer to the system. 

\section{Acknowledgments}
This work is part of a project that has received funding from the European Union’s Horizon 2020 research and innovation programme under the Marie Skłodowska-Curie grant agreement No 765452.  The information and views set out in this publication are those of the author(s) and do not necessarily reflect the official opinion of the European Union. Neither the European Union institutions and bodies nor any person acting on their behalf may be held responsible for the use which may be made of the information contained therein. 

\bibliographystyle{ACM-Reference-Format}
\bibliography{bib}

\end{document}